\documentclass[twocolumn]{aastex7}
\usepackage{graphicx} 
\usepackage{amsmath}
\usepackage{tabularx}
\usepackage{indentfirst}

\usepackage{colortbl}

\begin{document}

\title{Accounting for Transit Timing Detectability: Biases in Planetary Radius and Orbital Period} 
\author[0000-0001-5592-6220]{Skylar D'Angiolillo}
\affiliation{Department of Physics, The College of New Jersey, 2000 Pennington Road, Ewing, NJ 08628, USA}
\email{skylar.dangiolillo@email.ucr.edu}
\author[0000-0003-3750-0183]{Daniel Fabrycky}
\affiliation{Department of Astronomy \& Astrophysics, The University of Chicago, 5640 South Ellis Avenue, Chicago, IL 60637, USA}
\email{fabrycky@uchicago.edu}
\author[0000-0003-2372-1364]{Mariah G. MacDonald}
\affiliation{Department of Physics, The College of New Jersey, 2000 Pennington Road, Ewing, NJ 08628, USA}
\email{macdonam@tcnj.edu}

\begin{abstract}
    Transit Timing Variations (TTVs) are deviations from the time an observer would expect to see an exoplanet transit its host star. In multi-planetary systems, significant TTVs may indicate the presence of another body in the system gravitationally interacting with the transiting exoplanet. \citet{2016Holczer} catalogs 2599 Kepler Objects of Interest (KOIs) and provides a  statistical analysis of their TTVs for candidates that have transited at least seven times. However, this conservative limit on the number of transits neglects long-period KOIs. Therefore, we extend the statistical analysis performed by \cite{2016Holczer} to the population of KOIs that have between three and six transits. We identify six KOIs, three of which have Kepler names (Kepler-103 c, Kepler-90 g, and Kepler-1662 c), with significant TTV signals that were originally overlooked by \cite{2016Holczer}. Additionally, we search for trends regarding the planetary radius and orbital period of KOIs with significant TTVs. 
    Through a survival analysis, we determine that planets with shorter orbital periods require larger TTV signals to be detected compared to longer period planets, regardless of planetary size. Uncovering trends in TTV statistics such as this one will provide targets for future forward modeling of planetary architectures.
\end{abstract}

\section{Introduction}

\subsection{Transit timing variations}

On the heels of the first detection of a transiting planet \citep{2000Charbonneau, 2000Henry} came the first constraints on its dynamics. Due to the lack of transit timing variations, limits were placed on other bodies in the system, notably satellites \citep{2001Brown}. While there were still only a handful of confirmed transiting planets at the time, \cite{2005Agol} and \cite{2005Holman} fully fleshed out the size of timing variations to be expected by perturbing planets of various orbits. They found that perturbers, especially those with orbital periods resonant with the known planet, could be easily detected. On that basis, the first timing dataset with $\sim10$ transits was shown to lack Earth-mass planets in resonant orbits \citep{2005Steffen}.

Along with NASA's Kepler mission came an embarrassment of riches of transiting planets with variable periods \citep{2011Ford}; however, very few of the transit timing variations (TTVs) could be interpreted \citep{2010Holman, 2011Lissauer}. The variations of some of these signals were resolved as more data came in \citep{2012Carter, 2014Nesvorny, 2016Mills}, but many remain with the orbital parameters of the perturber yet-unsolved today. A thorough analysis of the transit timings of the complete long-cadence dataset was performed by \cite{2016Holczer}\defcitealias{2016Holczer}{H16}\citepalias{2016Holczer} (see also \citealt{2015Rowe,2018Ofir}). The detection of TTVs in 260 planet candidates has captured most investigators' attention. 

\subsection{Previous work on TTV statistics} 

Although \citetalias{2016Holczer} provided descriptive statistics, they did not search for trends in TTV signal with respect to planet properties. \cite{2014Xie} studied systems with single transiting planets and multiple transiting planets and found the latter had statistically higher TTV fractions. This finding is natural if the single transiting planets are truly the only planets orbiting their stars, or if the other planets in those systems are far enough away to perturb the transiting planet less than in systems where multiple planets transit. Additionally, analysis of the Kepler survey performed by \cite{2011Latham} and \cite{2012Steffen} found that large planets are more typically single (though this conclusion was moderated recently; \citealt{2023Wu}). Putting these effects together may predict that large planets will have detectable TTVs less often. 

\citet{2005Agol} theoretically predict that the amplitude and timescale of TTV signals are sensitive functions of planetary arrangement, especially period ratio. While we do not test the scaling presented by \citet{2005Agol}, we analyze observational trends in TTV detectability across orbital periods. Furthermore, TTVs are influenced by dynamical factors such as resonance. Planets near resonance cause a large amplitude and long timescale variation to a planet's time of transit. Because of this, if resonances are preferentially more likely to occur among larger planets  \citep{2015Winn}, then larger planets should be more likely to have TTVs than smaller planets. This trend differs from that concluded in the previous paragraph. 

Here, we depart from these studies by asking how the non-detections can be understood. Each upper-limit has a role to play in determining the frequency of TTVs of various amplitudes and timescales. There are many more non-detections than detections, and because the precision of timing data differs widely from object-to-object, we must pursue a statistical survival analysis \citep{1966Mantel} to make sense of any trends. 

\subsection{Expectations of trends}

\cite{2015Wang} studied long-period transiting planets from the Kepler population and found that more than half of exoplanets with a minimum of three transits exhibit TTVs. This study suggests that there are additional components that dynamically interact with these transiting planet candidates \citep{2015Wang}.  
This surprising discovery was tested for somewhat shorter periods by \cite{2019Martin}, who found that 15 out of 40 long-period planets (P= 100-200 days) with precise TTV data had statistically significant timing variations. 

Gas giants, which preferentially lie at $\sim 1$~AU and beyond, could generate large perturbations \citep{2008Cumming}. Since transit timing variations are a strong function of period ratio \citep{2005Agol}, it seems that the orbital period of strongly perturbed planets should be at least several hundred days.
A recent statistical study of transit duration variations (TDV) \citep{2021Shahaf,2021Millholland} and the survival analysis of \cite{2021Shahaf} 
show that longer orbital-period planets are torqued more from their orbital planes than closer-in planets. Gas-giant planets could be responsible for some of these TDVs, as they are thought to have inclinations of $\sim 10^\circ$ relative to the inner planets \citep{2020Masuda}.

However, approximately 10 times more cases of robust TTVs than TDVs have been found \citep{2016Holczer, 2021Shahaf}. Therefore, we seek to perform a survival analysis on these data to determine if the trends surrounding orbital periods are correct and significant. It is possible that planets are similarly perturbed regardless of their period and that it is easier to measure these perturbations at longer orbital periods. Accordingly, the null hypothesis would be that of \cite{2012Tremaine}, who postulate that the statistics of planetary systems is separable. In this case, if the population of systems has no intrinsic scale such that the angular dynamical interactions are the same regardless of period, then the probability of one planet having a large TTV signal (normalized to its orbital period) should be independent of period. Radial velocity surveys \citep{2008Cumming, 2019Fernandes} have already shown preferential scales for the gas-giant population. Similarly, we seek evidence for the break-down in separability in TTV statistics.

If the separability hypothesis holds, then for most planetary architectures, the TTV signal does not depend on the planetary size. One exception is that a truly resonant TTV signal can depend on the mass of the transiting planet \citep{2016Nesvorny, 2017SaadOlivera}– but these systems are rare. We suspect that separability does not hold in non-resonant cases, as found by ``peas in a pod’’ correlations \citep{2017Millholland,2018Weiss}, although these results have their detractors \citep{2020Zhu, 2020Murchikova}.

In this work, we aim to explore if the detectability of planets, in terms of TTV occurrence and strength, is a function of orbital period or planetary radius. The \cite{2016Holczer} dataset consists of many detections with a much larger number of upper limits. Here we chronicle our methods to serve as a guide to the difficulties of answering these questions.

\subsection{Outline}

In Section~\ref{sec: methods} we employ TTV data from \cite{2016Holczer} and recreate the statistical analysis used in the previous assessment of \cite{2016Holczer}, performed by \cite{2019Martin}, and compare these methods with other evaluations. We further outline the rationale behind the data selection for our analysis, discuss the statistical background of our survival analysis, and describe how we determine our threshold for significance. More specifically, we use False Discovery Rate statistics to determine the p-values at which the TTVs of different systems should be considered significant, using the method described in \cite{1995BenjaminiHochberg} and defaulting to a False Discovery Rate of 0.01 (1\%). Departing from \citetalias{2016Holczer}, we allow for analysis of systems with as few as three transits, compared to a minimum of seven transits in \citetalias{2016Holczer}. This leads to our discussion, in Section~\ref{sec: six_sig}, of six new significant TTV signals that our methods for detecting long-period KOIs uncover. Three of these systems were not analyzed by \citetalias{2016Holczer}, but were later verified by other means, and the other three systems are good candidates for further observation. In Section~\ref{sec: randp}, we continue our data interpretation, regarding the distribution of TTV amplitudes with respect to planetary radius and orbital period, through the methods in the prior sections. Specifically, we implement Survival Analysis statistics to compare the occurrence of significant TTVs in short- versus long-period systems, and small versus large radius planets, finding that shorter-period planets require larger TTVs to be detected, regardless of planetary size. Section~\ref{sec: con} concludes with interpretations of our findings.

\section{Data Selection and Methodology}\label{sec: methods}

Here we describe our two separate analyses, their data inputs, and our statistical methods. We discuss and recreate the findings of \cite{2019Martin} based on the statistical methods that \cite{2016Holczer} use to examine long-period KOIs in Section~\ref{sec:longKOI}. In Section~\ref{sec:data_select}, we expand upon these methods by broadening the conservative scope of the data selection criteria used by \cite{2016Holczer}. In Sections~\ref{sec: survival_analysis} and \ref{sec:sig_determ}, we describe the methodology behind our second analysis performed in Section~\ref{sec: randp},  outlining the statistical background of a survival analysis before employing False Discovery Rate Statistics to set a more accurate significance level on likely detections.

\subsection{TTV Signals of Long-period KOIs} \label{sec:longKOI}
We use transit timing variation measurements from \cite{2016Holczer} to analyze KOIs with orbital periods greater than 100 days \((P > 100)\), which we call ``long-period KOIs,'' and replicate the results of \citetalias{2016Holczer}. We calculate the O-C uncertainty median (\(\sigma _{TT}\)), O-C scatter (\(s_{O-C}\)), and p-value of the scatter divided by the uncertainty (\(p-s/\sigma\)) using the O-C and O-C uncertainty from Table 3 \citepalias{2016Holczer} and the orbital period from Table 2 \citepalias{2016Holczer}\footnote{ All lines in Table 3 \citepalias{2016Holczer} containing an asterisk were removed prior to the calculations}. We calculate the O-C scatter by multiplying the median absolute deviation by 1.4826, and we determined the p-value from \(\chi ^{2}_{mod}\):

\begin{equation}
   \chi^{2}_{mod} = n \times(s_{O-C}/ ( 1.48 \times \sigma _{TT}))^{2}
   \label{eq:mod_chi_sq}
\end{equation}

\noindent where $\sigma _{TT}$ is the median uncertainty and $n$ is the number of measurements.

In their discussion of Kepler-1625 b, \cite{2019Martin} analyze the planet's TTVs. They find that the normalized scatter of the TTVs for this planet is equal to $2.40 \times 10^{-5}$ and that the normalized median error is $1.55\times 10^{-5}$. \cite{2019Martin} then extends this analysis to the long-period KOIs included in \citetalias{2016Holczer}.

We recreate the findings of \cite{2019Martin}, who note that there exist 40 planets for which the median uncertainty normalized by the orbital period is less than or equal to \(1.55 \times 10^{-5}\) (i.e., \(\sigma_{TT} /P\le  1.55\times 10^{-5})\) and have a larger TTV scatter  \([s_{O-C} /P]> 2.40 \times 10^{-5}\). Of those 40 planets, 15 are also significant with a \(log (p)<-8.8\) \citep{2019Martin}. As \cite{2019Martin} assigned this \(log (p)<-8.8\) significance level based on the most conservative estimate for the given data, we are interested in determining if additional planets could be deemed significant if we employ a more appropriate method for establishing this cut-off. Additionally, more planets may meet the criteria for a significant detection if the conservative limit of at least seven transits that \citetalias{2016Holczer} places on transit number is amended. Therefore, in the following section, we employ an empirical distribution function to compare long-period KOIs from the \citetalias{2016Holczer} catalog with significant TTV signals to non-detections. This empirical distribution function describes the cumulative distribution of these observations with our discrete variable corresponding to a significant p-value.

\subsection{Less Conservative Data Selection Criteria} \label{sec:data_select}
\cite{2016Holczer} analyzed TTVs for KOIs within the scope of certain limitations. One such limitation is that the KOI must have at least seven transits to be analyzed in this publication \citepalias{2016Holczer}. However, this cut-off potentially excludes transiting planets with long orbital periods as they would have fewer opportunities to transit than short period planets during the finite duration of the Kepler survey. 

Therefore, we use transit timing variation measurements from Table 3 of \citetalias{2016Holczer} to analyze KOIs with orbital periods greater than 100 days \((P > 100)\) that have between three and six transits. 
We find 104 KOIs which met this criteria, excluding known false positives. Of these 104 KOIs, six have statistically significant p-values, indicating a detection, which we discuss further below. To produce upper limits on detections, we remove these six KOIs. We then use the median of the O-C error to compute the median TTV uncertainty for the remaining 98 KOIs. We use this median uncertainty to derive the critical O-C scatter value for each KOI (Eq.~\ref{eq:Scatter_critical}).

\begin{equation}
 s_{O-C critical}= \sigma _{TT} \times\sqrt{(\chi^{2}_{critical}/n)-1} 
 \label{eq:Scatter_critical}
\end{equation}

\noindent$s_{O-C critical}$ represents the threshold $s_{O-C}$ value at which the resulting $\chi^{2}$ p-value would be considered significant.\footnote{ Eq.~\ref{eq:Scatter_critical} does not use MAD-based scatter, so it does not require the 1.483 scaling factor.}

\subsection{Survival Analysis Statistics}\label{sec: survival_analysis}
\cite{1966Mantel} introduced survival analysis as a way to compare the effectiveness of medical treatment based on patient survival. \cite{1985Feigelson} illustrates how these survival analysis techniques can be applied to astronomical questions through the examination of six data sets. We have adapted this method to our work by equating a significant level of TTVs, in accordance with the False Discovery Rate, to the death of a patient. We adopt the notation of \cite{2006Machin} and reproduce their equations 3.1 and 3.2 below. 

Here, the expected number of significant detections for long- (A) and short- (B) period planets with TTV amplitudes greater than or equal to $t$ is 
\begin{equation}
    E_{At}= r_{t}m_{t}/ N_{t},
    E_{Bt}= r_{t}n_{t}/ N_{t}
\end{equation}

\noindent where $m_{t}$ represents the total number of long-period KOIs, $r_{t}$ is the total number of significant KOIs (both short- and long-period), $N_{t}$ is the total number of all KOIs included in the study, and the conditional variance, adapted from equation 3 of \cite{1966Mantel}, $V$ is 

\begin{equation}
 V_{A t}=\frac{m_{t}n_{t}r_{t}s_{t}}{N_{t}^2 (N_{t}-1)}
 \label{eq:Mantel_3}
\end{equation}
\noindent where $s_{t}$ is the total number of non-significant KOIs (both short- and long-period) and $n_{t}$ is the total number of short-period KOIs.

Therefore, once the expected number of significant detections is calculated for each amplitude, the number of observed and expected significant detections are summed to give

\begin{equation}
    O_{A}=\Sigma O_{At},  O_{B}=\Sigma O_{Bt},
     E_{A}=\Sigma E_{At},  
      E_{B}=\Sigma E_{Bt}
\end{equation}

The continuity-corrected chi-square \citep[equation 3.12 of][]{2006Machin}, which compares the probability of a KOI having statistically significant or non-significant TTVs depending on its orbital period, is
\begin{equation}
 \chi^2 = \frac{(O_{A} -E_{A})^2}{V}
 \label{eq:Mantel_1}
\end{equation}

\noindent where we sum the individual $V_{At}$ at each amplitude so that $V = \sum V_{At}$.

We perform a survival analysis on all KOIs with at least seven transits, comparing the strength of TTV signals based on orbital period and planetary radius. For each KOI, we compute the continuity-corrected chi-square statistic (Eq.~\ref{eq:Mantel_1}) to determine which of the TTVs are statistically significant.

We recreate equation 3.3 from \citet{2006Machin} to calculate the Logrank statistic below. 

\begin{equation}
    \chi^2_{Logrank}= \frac{(O_{A}-E_{A})^2}{E_{A}}+ \frac{(O_{B}-E_{B})^2}{E_{B}}
\end{equation}
\noindent We use this statistic to test the null hypothesis that the probability of detecting a significant TTV signal at any point in time is the same for both long-period and short-period planets.

\subsection{Determining Significance for Survival Analysis}\label{sec:sig_determ}

We wish to test a sample of systems for transit timing variations. After determining those systems with significant detections, we will need to correct the distribution of the sample for non-detections which may not have had enough sensitivity to detect variations at the same level. We also want our list of significant detections to exclude spurious detections that would improperly skew our distribution. 
This problem is one of multiple hypothesis testing. Therefore we use the procedure outlined by \cite{1995BenjaminiHochberg}\defcitealias{1995BenjaminiHochberg}{BH95}\citepalias{1995BenjaminiHochberg} which considers several factors: 
\begin{itemize}
    \item{ $p$-value -- the probability an individual sample will be falsely deemed significant, }
    \item{ $N$ -- number of trials total,}
    \item $k$ -- number of trials deemed significant,
    \item $\alpha$ -- false discovery rate, 
    \item $p_{\rm crit} = (k/N) \alpha$ -- \citetalias{1995BenjaminiHochberg} upper bound on the probability that the least significant of $k$ trials deemed significant, is actually a spurious detection,
    \item $\overline{FD} = k \alpha$ -- bound on the expected number of false detections, which is just the probability value of the least significant ``detection'' being a false detection, times the number of trials deemed significant. 
\end{itemize}
In order to control for type I errors, in which we would falsely reject the null hypothesis that there is not a significant level of TTV signal, we perform a false discovery rate (FDR) analysis. In accordance with \citetalias{1995BenjaminiHochberg}, we define the false discovery rate, $Q_{e}$ as \begin{equation}
 Q_{e} = E(Q )= E[V/(V+S)] = E(V/R)
 \label{eq:FDR}
\end{equation} where $V$ is the number of errors committed when a false positive is declared significant, $S$ is the number of true positives declared significant, and $R$ is the total number of discoveries deemed significant, or the total number of rejected null hypotheses \citepalias{1995BenjaminiHochberg}. Therefore, the false discovery rate is the expected number of false discoveries divided by the total number of discoveries deemed significant.

We aim to minimize the number of false detections, so we set our $\alpha$ value to 0.01, which means that there is only a 1\% chance that any individual detection is false. Because we have 225 significant detections, $\overline{FD}=k \alpha=2.25$. This results in us most likely having $2$--$3$ false detections out of our 225 significant detections. Since there is only a 1\% chance that any of detections we deem to be significant are actually not significant, we call this the ``likely detection list'' in each subset of the data, e.g. in a box of ($P$, $R_p$) space. For a more expansive list of TTV candidates, which necessarily have more false positives, we choose $\overline{FD}=0.05 k$ such that each item on our likely detection list has a false positive detection probability of less than 5\%.
That is, this should be considered 2-$\sigma$ evidence of detection of individual significant TTV, i.e. in need of follow-up data to confirm.

\begin{figure}[t!h]
    \centering
    \includegraphics[width=\linewidth] {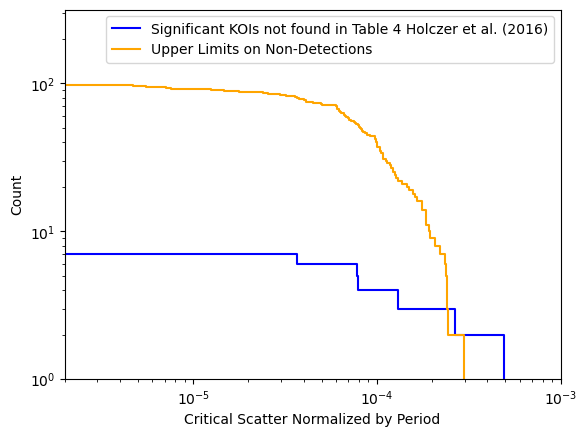}
    \caption{Cumulative distribution functions of the critical O-C scatter (from Eq.~\ref{eq:Scatter_critical}) for the 98 long-period KOIs excluded by \citetalias{2016Holczer} (orange) and of the O-C scatter for six statistically significant planet candidates (blue). These six candidates (KOIs: 108.02, 351.02, 868.01, 1209.01, 1477.01, and 1783.02) were not considered detections by \citetalias{2016Holczer} because they had fewer than seven transits. These KOIs meet the criteria for significance established by \citet{2019Martin} (i.e. $log(p)<-8.8$). Each step ``up'' in the blue curve, moving from right to left, indicates a significant signal within the step range, while the orange curve represents the cumulative count distribution of the 98 upper limits for the scatter with insignificant levels of TTV. We normalize both the critical and the measured O-C scatter by taking the root-mean square of the TTV divided by the best-fitting orbital period. Note that both axes are logarithmically scaled.}  
    \label{fig:Critial_Scatter}
\end{figure}

\section{Six New Statistically Significant TTV Signals}\label{sec: six_sig}

For each of the long-period KOIs included in Table 3 of \citetalias{2016Holczer} with three to six transits, we calculate the normalized TTV scatter and the associated $\chi^2_{mod}$ as described in Sections~\ref{sec:longKOI} and \ref{sec:data_select} above. From this $\chi^2_{mod}$, we recover an associated p-value, the probability of a random signal having $\chi^2_{mod}$ at least as extreme as the KOI's. A small p-value indicates a low probability of a Type I error, suggesting a statistically significant level of TTV scatter. More specifically, these KOIs meet the criteria established by \citet{2019Martin}, as they were identified during our recreation of their findings after relaxing the minimum transit number requirement (Sec.~\ref{sec:longKOI}). The p-values corresponding to each of these six detections are $log(p)<-13$.

We show the critical O-C scatter of non-detections, as described above in Section~\ref{sec:data_select}, along with the O-C scatter for these six significant KOIs in Figure~\ref{fig:Critial_Scatter}.
We normalize both sets of O-C scatter by the planet's orbital period found in Table 2 of \citetalias{2016Holczer}, thus ensuring that the length of the period does not skew the data and that the trends we see are not confounded by the orbital period. 
Using our less conservative requirement of three transits, we recover six KOIs with significant TTVs that were excluded from Table 4 of \citetalias{2016Holczer}---KOIs 108.02, 351.02, 868.01, 1209.01, 1477.01, and 1783.02---which we briefly discuss in turn below.

\begin{figure*}[t]
    \centering
    \includegraphics[width= 3.38in] {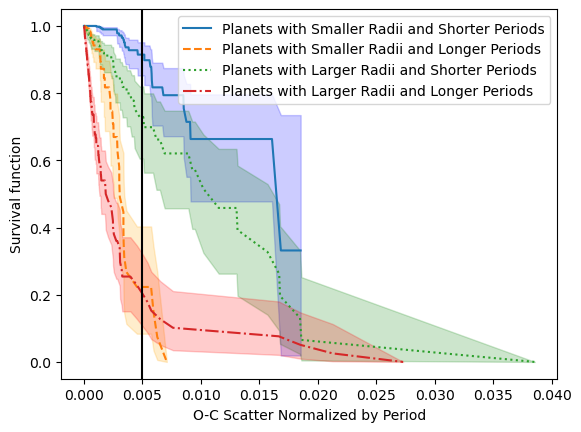}
    \includegraphics[width= 3.5in]{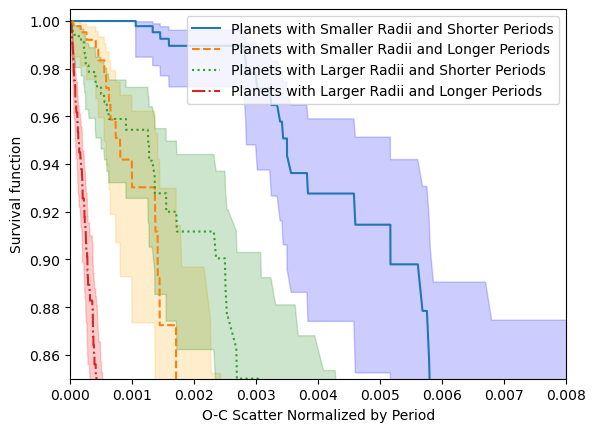}
    
    \caption{ Left) Kaplan-Meier graph depicting the proportion of KOIs for which we did not detect significant levels of TTV (y-axis) given the O-C scatter normalized by orbital period (x-axis) for the 2339 KOIs included in Table 4 of \citetalias{2016Holczer}. We group KOIs based on median planetary radius (2.482 Earth radii) and median orbital period (17.79 days), creating four groups: ``small radii / short period'' (690 KOIs, blue, solid line), ``small radii / long period'' (480 KOIs, orange, dashed line), ``large radii / short period'' (480 KOIs, green, dotted line), and ``large radii / long period'' (689 KOIs, red, dotted-dashed line). The shaded regions represent the 95\% confidence intervals for each survival function. There is a vertical black solid line denoting a normalized scatter of 0.005.  Right) Close-up of the survival function depicted on left between 0.85 and 1. The survival function dictates what percentage of KOIs in the group remain undetected have at least that amount of normalized O-C scatter. For example, $\sim$90\% of large planets with short periods (green, dotted) are not detected at a normalized O-C scatter of 0.0025 and $\sim$91\% of small planets with short periods (blue, solid) are not detected with a normalized O-C scatter of 0.005. Regardless of planet size, KOIs with short orbital periods require higher levels of TTVs to be detected. 
   }
    \label{fig: Kaplan-Meier}
\end{figure*}

\textbf{KOI-108.02:} Kepler-103 is a G-type star with two known transiting planets, with radii of $R_b=3.49^{+0.057}_{-0.054}~R_{\oplus}$ and $R_c=5.45\pm0.18~R_{\oplus}$ and orbital periods of $P_b=15.97$ and  $P_c=179.6$ days \citep{Bonomo2023}. The signal we detect here is that of Kepler-103 c. Both planets were confirmed through low false-positive probabilities based on a lack of nearby stellar companions, asteroseismetric analyses, and RV follow-up \citep{Marcy2014} prior to the analyses of \citetalias{2016Holczer}. 

\textbf{KOI-351.02:} Kepler-90 is a G-type star hosting eight transiting exoplanets, with radii ranging from $R_p=1.4~R_{\oplus}$ \citep[05/c, ][]{Weiss2024} to $R_p=1.0~R_J$ \citep[01/h, ][]{Weiss2024} and orbital periods from 7.0 to 331.6 days. Seven of these planets were confirmed by \citet{Cabrera2014} using TTV fitting, while Kepler-90 i was uniquely discovered through the use of neural nets \citep{Shallue2018}. Years later, \citet{Liang2021} discovered and fit the TTV signal between Kepler-90g (KOI-351.02) and Kepler-90h (KOI-351.01), the same signal we detect here, and report a low mass for Kepler-90g of $15.0^{+0.9}_{-0.8}~M_{\oplus}$, corresponding to a low bulk density of $0.15$ g cm$^{-3}.$ More recently, \citet{Weiss2024} constrained the masses of planets g and h using archival RV data.

\textbf{KOI-1783.02:} Kepler-1662 is a G-type star with two transiting gas giants. These planets, have orbital periods of 134.5 days and 284.2 days and radii of $R_b=8.86^{+0.25}_{-0.24}~R_{\oplus}$ and $R_c=5.44^{+0.52}_{-0.30}~R_{\oplus}$. \citet{Vissapragada2020} summarized prior work, including detection of the transit timing variations by \cite{2015Kipping} and \cite{2018Ofir}, and carried out dynamical fits to confirm KOI-1783.02 as a planet.

The three remaining KOIs with significant TTVs are KOI-868.01, KOI-1209.01, and KOI-1477.01. These three KOIs all remain the sole planet candidate in their respective systems, suggesting that these candidates could be planets and they could live in multi-planet systems. Beyond \citetalias{2016Holczer}, no study has considered these systems, but confirming the planetary status of these KOIs is beyond the scope of this work. 

These six long-period KOIs were not analyzed by \citet{2016Holczer} because they contain fewer than seven transits. We therefore urge future investigations to implement similar statistical analyses as described here and to explore all candidates with three or more transits to avoid biasing against long-period planets.

\section{Planetary Radius and Orbital Period Comparisons}\label{sec: randp}

We perform a survival analysis 
on the data from Table 4 of \citetalias{2016Holczer}. This survival analysis consists of using False Discovery Rate (FDR) Statistics to more appropriately define the the p-value at which TTVs are considered to be significant. Since we are looking at many samples, it would be inappropriate to use a single p-value to determine a significant level of detection. This methodology provides a way to control the expected percentage of false discoveries \citep{2021Shahaf}. While the conventional procedure bases a significant discovery off of a predetermined p-value, the FDR method is proven by \cite{1995BenjaminiHochberg} to  better manage the purity of the sample. Similar to \citet{2021Shahaf}, we conclude our analysis by implementing Kaplan-Meier statistics to inform the detectability of KOIs based on their $O-C_{Scatter}$. For this analysis, we use the \texttt{life-lines} package \citep{DavidsonPilon2021}. Additionally, we use Log rank tests, and ad hoc analysis including a Student t-test, to follow-up comparisons between different planet populations.

\subsection{Trends in Exoplanet Detectability}
Table 4 of \citetalias{2016Holczer} details various calculations relating to the significance of long-term TTVs for 2339 KOIs with more than six transits. We assess the effect of planetary radius and orbital period on detectability of these 2339 KOIs using the p-value, period, and O-C scatter from Table 4 of \citetalias{2016Holczer} and the radius from \citet{2023Lissauer}. Through the use of false discovery rate statistics, outlined above in Section~\ref{sec: survival_analysis}, and using our false discovery rate of $\alpha = 1\%$, we find 225 of the 2339 KOIs to be significant at $p_{\rm crit}= 0.000962$. This is fewer than the 274 significant TTV systems (out of 2599) reported by \cite{2016Holczer}, five of which may be due to stellar activity. These differences arise mainly from differences in methodology: while \citetalias{2016Holczer} combined $\chi^{2}$ scatter analysis, power spectrum periodograms, alarm scores, and polynomial fitting to identify long-term TTVs, we apply a false discovery rate procedure based on \cite{1995BenjaminiHochberg} to control for false positives, yielding a more conservative detection list. Additionally, our analysis is limited to the 2399 KOIs included in Table 4 of \citetalias{2016Holczer},  which reduces our sample size and likely contributes to the lower number of significant TTVs.

We split the 2339 KOIs in this dataset into four groups based on their orbital period and radius.  
We consider KOIs with radii less than or equal to the median (2.482 $R_{\oplus}$) as ``small'', while those with radii greater than the median were considered to be ``large''. Similarly, KOIs with orbital periods less than or equal to the median (17.79 days) were considered to be ``short'', while those with periods greater than the median were considered to be ``long''. Thus, we construct four groups: ``small radii / short period'' (690 KOIs), ``small radii / long period'' (480 KOIs), ``large radii / short period'' (480 KOIs), and ``large radii / long period'' (689 KOIs). 

We compare each group against each other via Kaplan - Meier statistics (Figure~\ref{fig: Kaplan-Meier}) and log rank tests (Table~\ref{table:Logrank}) which incorporate upper limits. We estimate that only about 20\% of large planets with long orbital periods (red dot-dashed curve in \ref{fig: Kaplan-Meier}) remain undetected with a normalized O-C scatter of 0.005 (vertical black line), while 73\% of large planets with short orbital periods(green dotted curve) are considered non-detections with normalized O-C scatter of 0.005. Similarly, approximately 22\% of small planets with long orbital periods (orange dashed curve) that have normalized O-C scatter of 0.005 are not detected, while 91\% of their short period counterparts (blue solid curve) with a normalized O-C scatter of 0.005 are not detected. In other words, only $\sim$27\% of large planets with short orbital periods and $\sim$9\% of small planets with short orbital periods are detected at the same level of normalized scatter that $\sim$80\% of large planets with long orbital periods and $\sim$78\% of small planets with long orbital periods are detected. This trend continues across higher levels of scatter as KOIs with longer orbital periods are detected more readily than KOIs with shorter periods. 

\begin{figure*}[t]
   \centering
    \includegraphics[width=0.7\textwidth] {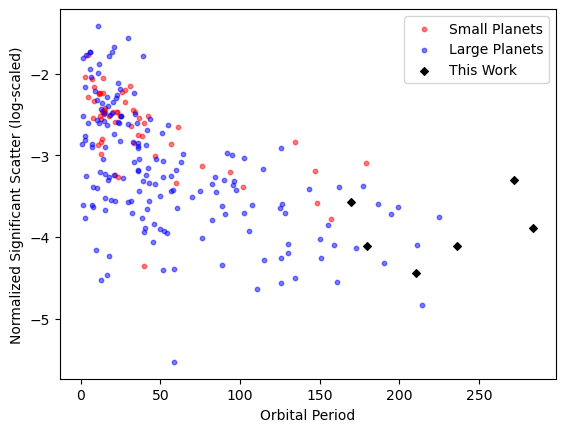}
    \caption{Scatter plot visualizing the distribution of normalized O-C scatter (log-scaled, y-axis) for significant KOIs with respect to their orbital period in days (x-axis). We group KOIs based on median planetary radius (2.482 Earth radii) with small planets represented by red circles and large planets represented by blue circles. Additionally, we include the six significant KOIs that we identify in Section~\ref{sec:sig_determ} as black diamonds. 
    We find that short period planets (those with P$\lesssim18$ days) exhibit a larger range of normalized O-C scatter than their long period counterparts. However, the lower envelope of the distribution ($\approx3\times10^{-5}$) is consistent across all periods. This reflects an observational bias, with short-period planets requiring stronger TTV signals to be detected, rather than an intrinsic difference in TTV amplitude across orbital periods. Although the wider range of O-C scatter at short orbital periods is primarily driven by large planets, which span several orders of magnitude, small planets also exhibit substantial normalized TTV amplitudes, despite having a narrower range.}
    \label{fig: Scatter_plt}
\end{figure*}

Each subgroup is formally significant from one another, however, the period differences are much more interesting and the major result of this paper. These results are also visualized in Figure~\ref{fig: Scatter_plt} which depicts the distribution of the normalized scatter for significant KOIs with respect to their orbital periods. 
This graph further supports the conclusion that short-period KOIs require higher levels of TTV to be detected.

\begin{deluxetable}{ccccc}
\tabletypesize{\footnotesize}
\tablecolumns{5}
\tablewidth{0pt}
\tablecaption{ Planetary Radius and Orbital Period Log Rank Test Results \label{table:Logrank}}
\tablehead{
\colhead{} & \colhead{ Small $R_{p}$ / Short $P$} & \colhead{Small $R_{p}$ / Long $P$} & \colhead{Large $R_{p}$ / Short $P$ } & \colhead{Large $R_{p}$ / Long $P$}}
\startdata
Small Radii / Short Period & \nodata & 114.08 & 23.43 & 222.06 \\ 
 Small Radii / Long Period & 114.08 & \nodata & 14.30 & 33.84\\
Large Radii / Short Period  & 23.43 & 14.30 & \nodata & 73.43 \\
Large Radii / Long Period  & 222.06  & 33.84  & 73.43 & \nodata \\
\enddata
\tablecomments{p-values (-log2(p)) for log rank tests between sub-populations in  Table 4 of \citetalias{2016Holczer}. All computed values are less than 0.005, indicating statistically significant differences in the samples.}
\end{deluxetable}

\subsection{Additional Analysis: ``Weeks''}

As short-period planets have more opportunities to transit their host star than planets with longer periods, we wish to ensure that our results are robust against the preference for short-period TTV.
That is, short-period planets have an opportunity to show variations on a longer dynamical timescale (relative to their orbital periods), because as seen in Figure 6 of \citet{2013Mazeh}, the majority of TTV detections have TTV timescales at least an order of magnitude longer than the orbital period. It is possible, then, that those long timescales are sufficiently sampled for the short orbital periods but not for the long orbital periods.

We introduce an additional analysis intended to counter this bias, which we term a ``Weeks Analysis'' due to privileging groups of seven transits. We look for whether TTVs are apparent in just seven transits which filters out variation on longer timescales. In this analysis, we extract the transit number, O-C, and O-C error values from Table 3 of \citetalias{2016Holczer}. As \citetalias{2016Holczer} only examined KOIs with at least seven transits, we consider this to be the minimum analyzable dataset. Therefore, we began our analysis by selecting  KOIs with at least seven transits. These first seven datapoints were considered to be the first ``week'' belonging to that KOI. Similarly,  data from transits numbered 8--14 were considered to correspond to the second ``week'' and data corresponding to transits 15--21 belong to the third ``week''. For long-period KOIs ($P \geq 65$ days), we analyzed the first ``weeks'' worth of data, for medium-period KOIs ($50 < P < 65$ days) we analyzed the first two ``weeks'', and for short period KOIs ($P \leq 50$ days) we analyzed all three ``weeks''. We did not analyze more than the first twenty-one transits for any KOI. We then perform a similar survival analysis and False Discovery Rate Statistics analysis on these alternately analyzed data, as described in Section~\ref{sec:sig_determ}, to determine a significance level ($P_{crit}= 0.00147$, $\alpha=1\%$). This resulted in a total of 6128 alternately analyzed KOIs. 699 out of 5538 short-period KOIs, 69 out of the 258 medium-period KOIs, and 132 out of the 332 KOIs were found to be significant at this level. 

\begin{figure*}[th]
    \centering
    \includegraphics[width= 0.7\textwidth] {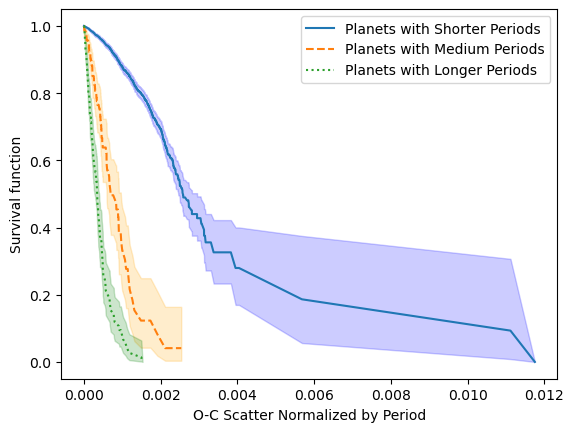}
    \caption{Kaplan-Meier plot depicting the data from Table 3 of \citetalias{2016Holczer} reconstructed into ``weeks.'' Here, we treat each set of seven consecutive transits as its own KOI to remove the bias of short period planets being observed more frequently. The blue solid function corresponds to short-period ``weeks'' ($P \leq 50$ days), the orange dashed function corresponds to medium-period ``weeks'' ($50 < P < 65$ days), and the green dotted function corresponds to long-period ``weeks'' ($P \geq 65$ days). 
The survival function dictates what percentage of KOIs in the group remain undetected at that level of normalized O-C scatter. The long tail of the short-period planet distribution highlights that 40\% of such planets are still undetected with a normalized O-C scatter of approximately 0.0031, whereas no planets with medium or long orbital periods require such extreme scatter to be detected. By exploring each KOI as a set of seven transits, we find that KOIs with inherently short orbital periods require more O-C scatter, and therefore greater TTVs, than long period planets to be detected.     }
    \label{fig: Weeks_Analysis}
\end{figure*}

Our results support the finding that planets with shorter periods require larger TTV signal to be detected, as depicted in Figure \ref{fig: Weeks_Analysis}. Additionally, we perform a log rank test on these data. While each group of ``weeks'' is significantly different from one another, the long-period group was the easiest to detect with the least amount of scatter (Table \ref{table:Weeks_Table}). While a $\sim$20\% rate of non-detection ($\sim$80\% of planets detected) is achieved with a normalized scatter of 0.0006 for planets with longer periods and 0.0012 for planets with medium periods, a normalized scatter of 0.0055 is required to reach the same level of detection for planets with shorter orbital periods. Therefore, we see that the trend surrounding KOIs with shorter orbital periods requiring greater levels of TTV still holds even when we correct for potential sampling bias.

\begin{deluxetable}{cccc} [th]
\tabletypesize{\footnotesize}
\tablecolumns{5}
\tablewidth{0pt}
\tablecaption{``Weeks'' Analysis Log Rank Test Results \label{table:Weeks_Table}}
\tablehead{
\colhead{} & \colhead{  Short Period} & \colhead{Medium Period} & \colhead{Long Period}} 
\startdata
Short Period & \nodata & 259.51 & inf \\ 
Medium Period & 259.51 & \nodata & 37.28 \\
Long Period  & inf & 37.28 & \nodata\\
\enddata
\tablecomments{p-values (-log2(p)) for log rank tests between long, medium, and short orbital period ``weeks''. All computed values are less than 0.005, indicating that the samples are statistically different.}
\end{deluxetable}

\subsection{Levels of Scatter}
Because we recover that short-period planets require larger TTV signal to be detected, we wish to test if any trends exist in the strength of TTV signals. We first group KOIs by the significance of their TTV signals. Within these two groups--KOIs with significant signals and KOIs with non-significant signals--we further group KOIs by radius and orbital period, as before. We then explore if the mean of each group is significantly different from the mean of the other groups. 

Using a Student t-test\footnote{The Student t-test assumes that the variance between groups is equal. Relaxing this assumption, by performing instead a Welch's t-test, leads to the same conclusions}, we test the null hypothesis that the means of the groups are equivalent. For KOIs with significant TTV signals, we fail to reject the null hypothesis ($p>$0.05) and must conclude that the amount of O-C scatter is consistent across groups of radius and orbital period. However, KOIs with non-significant TTV signals follow the expected trend for TTV amplitude: large planets have significantly less O-C scatter than small planets (t = -9.59, $p<10^{-16}$), and short-period planets have significantly less O-C scatter than long-period planets (t = -6.07, $p$ = 1.5e-9).

If we instead test if the amount of O-C scatter normalized by the orbital period differs between these different groups, we find similar results. For KOIs with significant levels of normalized O-C scatter, we fail to reject the null hypothesis ($p>$0.05) and conclude that the amount of normalized O-C scatter is consistent across groups of radius and orbital period. We again find the expected trends for KOIs with non-significant TTV signals, where large and short-period planets have significantly less normalized O-C scatter than small and long-period planets.

\begin{figure}
    \centering
    \includegraphics[width=\linewidth]{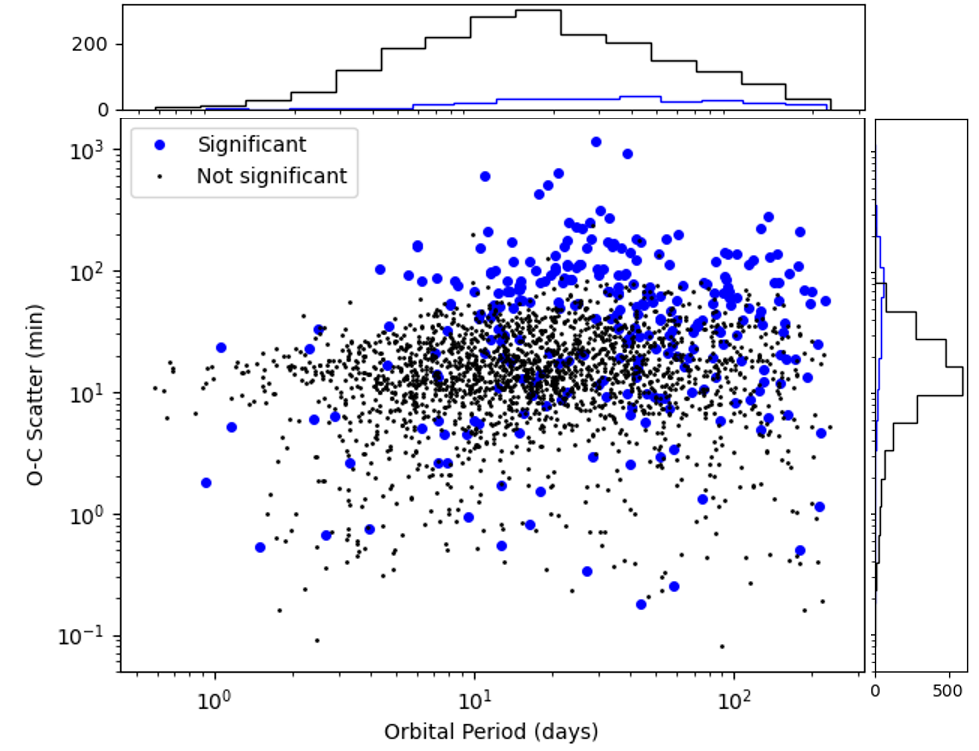}
    \includegraphics[width=\linewidth]{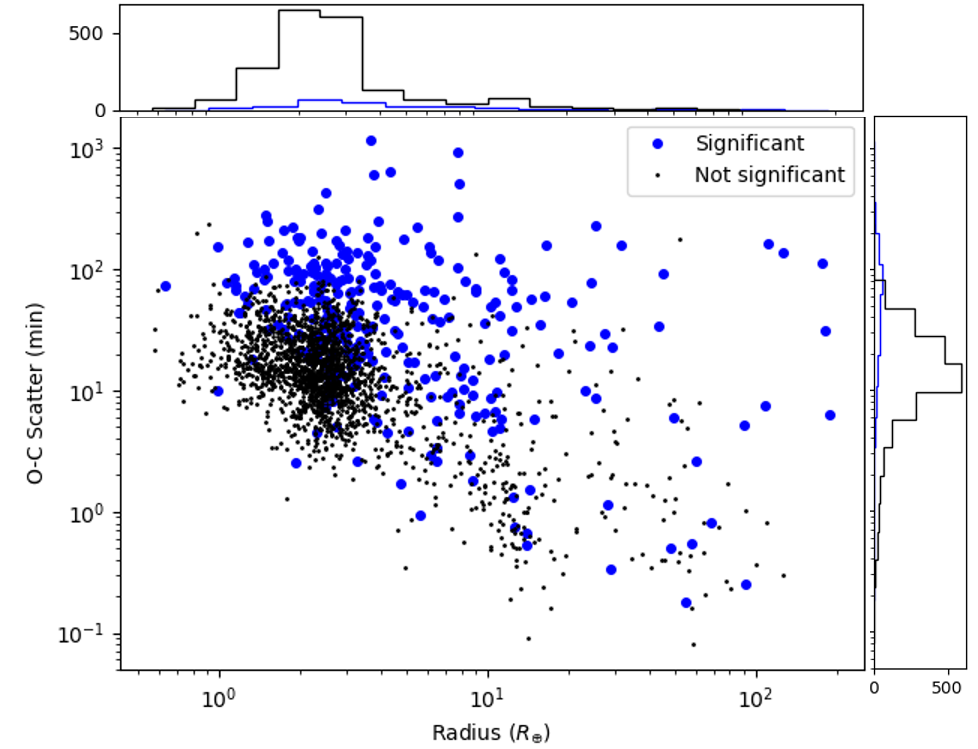}
    \caption{Top) O-C scatter in minutes vs. orbital period in days for KOIs with significant (blue) and non-significant (black) levels of O-C scatter. Bottom) O-C scatter in minutes vs. radius in $R_{\oplus}$.}
    \label{fig:O-CvsPR}
\end{figure}

To further explore why KOIs with significant TTV signals do not show variations in O-C scatter based on radius or orbital period, and KOIs with non-significant signals do show such variation, we show the distributions of O-C scatter as functions of radius and of orbital period for these two groups in Figure~\ref{fig:O-CvsPR}. We see that, while orbital period produces no noticeable trend in O-C scatter, there exists a trend in radius. For KOIs with non-significant O-C scatter, smaller planets tend to have more O-C scatter. We confirm this negative, nonlinear correlation using a $\tau$ test, a non-parametric hypothesis test for dependence ($\tau$ = -0.30, $p<10^{-16}$). Although this dependence exists for KOIs with significant TTV signals, it is weaker ($\tau$ = -0.25, $p$ = 5.8e-10). We also find that radius is negatively, weakly, and nonlinearly correlated to normalized O-C scatter for both KOIs with significant signals ($\tau$=-0.27, $p$=7e-12) and KOIs with non-significant O-C scatter ($\tau$=-0.34, $p<10^{-16}$).

Our statistical tests allow us to make two claims in support of our previous results. First, that our survival analysis recovered that only short-period KOIs with the highest levels of scatter (i.e. long-period KOI-like scatter) are determined to be significant, likely due to the existing detection bias. Second, that our discovery that short-period KOIs require higher levels of scatter, and that this requirement is \textit{robust} to radius, likely stems from the correlation between radius and O-C scatter. When we partition the KOIs based on the median radius and orbital period, the TTV signals of the short-period and small KOIs are overwhelmingly non-significant (96\%). Separately, the significant long-period KOIs are more likely to be small since we have detected relatively few large  KOIs at long orbital periods. Combining this paucity with the correlation between scatter and radius, we find that long-period KOIs are more likely to have large O-C scatter. Short-period KOIs are therefore likely to require larger O-C scatter to be considered significant.  

\section{Conclusions}
\label{sec: con}
While significant TTVs are known to potentially indicate the presence of another orbiting body in a multi-planetary system, there is little known about the trends surrounding this method of exoplanet detection. Therefore, we sought out upper limits for detectable exoplanets and investigated the impact of orbital period and planetary radius on the detectability of TTV. 

Following \citet{2016Holczer}, we search for significant levels of TTVs in all KOIs with at least three transits. We find that the conservative limit of seven transits placed by \citet{2016Holczer} excluded signals from planets with long periods, recovering six additional KOIs with significant levels of TTV, three of which remain planetary candidates (KOIs 868.01, 1209.01, and 1477.01). These planetary candidates should be studied for further validations, so that they may potentially be confirmed as exoplanets.

Our investigation employs False Discovery Rate statistics to determine the criteria for a significant exoplanet detection. Rather than using a common arbitrary p-value, such as $p= 0.01$, we set our false discovery rate to 1\% to minimize the possibility of incorrectly marking a detection as significant. Therefore, we set a rigorous standard for any significant detections, laying a solid statistical foundation for the rest of our analysis.

We then perform a survival analysis, which characterizes the probability of a KOI producing a significant level of TTV signal based on its radius and orbital period. We find that short-period planets require higher levels of normalized O-C scatter to be detected (Fig.~\ref{fig: Kaplan-Meier}), whereas long-period planets do not need as strong a TTV signal to be detected. We explore the amount of TTV signal as a function of both radius and orbital period, finding that the amount of O-C scatter does not significantly differ between groups of different orbital periods and radii for KOIs with significant TTV signals, but that both large planets and short-period planets exhibit less O-C scatter for KOIs with non-significant TTV signals. We further find that the amount of O-C scatter is negatively, nonlinearly correlated to the planet's radius, for both KOIs with and without significant levels of TTV, meaning that the amount of O-C scatter tends to decrease with an increase in planet radius. This correlation might explain why short-period planets require higher levels of O-C scatter to be considered significant, regardless of planet radius.  

While we do see strong TTV signals in KOIs with longer periods, $O-C_{Scatter} > 0.020$ (Fig.~\ref{fig: Kaplan-Meier}), KOIs with shorter periods are not detected as often even when they have the same level of period-normalized TTV strength. The mathematical analysis of the motion of a three-body system, given by \cite{2005Agol}, shows that the strongest TTV signals should be observed when the transiting planet has a long period. There are various reasons as to why this trend in planetary detectability could exist. Gas giants preferentially occupying a distance of $\sim1$ AU from their host star \citep{2008Cumming}, thus belonging to the long-period group, may be a factor. Additionally, the amplitude of TTV signals increases with the eccentricity of both the transiting and perturbing planets \citep{2009Nesvorný}. As long-period planets are more often on eccentric orbits, this may also contribute to the increased detectability of long-period planets.

Furthermore, Hot Jupiters are ideal planetary targets for short-period observations. However, it is unlikely to detect a planetary companion of a Hot Jupiter using the TTV method as small companion mass, as well as large period ratios, all hinder detection \citep{2012Steffen}. Furthermore, the signals of short-period transits yield few data points per each transit event \citep{2018Ofir}. Without the wealth of data that long-period transits produce, variations in short-period transit timing may not be captured or may fail to meet the criteria for a significant detection. Therefore, it is plausible that the trend we observe in the \citetalias{2016Holczer} data is a result of detection bias against weaker TTV signals from short-period planets, allowing only the strongest signals to be considered significant detections.

Overall, we hope that our work may serve as a model for future searches into trends surrounding the detection of exoplanets through the TTV method. Specifically, the implementation of survival analyses and False Discovery Rate statistics more accurately predict and constrain significant levels of TTV signals. Additionally, we recommend including KOIs with even a few transits in future analyses to avoid biasing against long-period planets. We advocate for future investigations of TTV statistics based on forward modeling, which would naturally fold in the period trend that we have uncovered. 

\begin{acknowledgements}
    All the data used in this paper produced by the Kepler Mission can be found in MAST: \dataset[10.17909/T9059R]{http://dx.doi.org/10.17909/T9059R}. We thank the anonymous referee for their constructive comments which improved this manuscript.
\end{acknowledgements}

\bibliography{bib}
\bibliographystyle{aasjournal}

\end{document}